\documentclass{raa_twocolumn}            

\usepackage{graphicx,amsmath,amssymb}             

\begin{document}

   \title{A note on the anti-glitch of magnetar SGR 1935+2154
}

   \volnopage{Vol.0 (20xx) No.0, 000--000}      
   \setcounter{page}{1}          

   \author{H. Tong}

   \institute{School of Physics and Materials Science, Guangzhou University, Guangzhou 510006, China;
   {\it tonghao@gzhu.edu.cn}\\
   }

   \date{Received~~2009 month day; accepted~~2009~~month day}

\abstract{The magnetar SGR 1935+2154 is reported to have an anti-glitch, accompanied by fast radio bursts, and transient pulsed radio emission. In the wind braking model, this triplet event tells people that
(1) SGR 1935+2154 does not have a strong particle wind and can be approximated by magnetic dipole braking in the persistent state; (2) Its anti-glitch is due to an enhanced particle wind, similar to the first anti-glitch in magnetars; (3) Its transient pulsed radio emission may be due to a decreasing emission beam during the outburst; (4) The enhanced particle acceleration potential and pulsar death line may not be the dominate factor.
}

   \authorrunning{H. Tong}            
   \titlerunning{On the anti-glitch of magnetar SGR 1935+2154}  

   \maketitle

%
%
\section{Introduction}

The magnetar SGR 1935+2154 is reported to have an anti-glitch (i.e., spin-down event) (Younes et al. 2022). The anti-glitch is also followed by fast radio bursts and pulsed radio emission (Younes et al. 2022; Good et al. 2020; Zhu et al. 2020). This will make SGR 1935+2154 the sixth magnetar to show pulsed radio emission (Camilo et al. 2006; Huang et al. 2021). At the same time, SGR 1935+2154 is also the first Galactic magnetar to show a fast radio burst (CHIME/FRB collaboration et al. 2020; Bochenek et al. 2020). This triplet of anti-glitch, fast radio bursts and pulsed radio emission may shield light on the physics of magnetar magnetosphere (Younes et al. 2022).

Previously, the magnetar spin-down behavior is proposed to be dominated by wind braking (Harding et al. 1999; Tong et al. 2013). When the anti-glitch was firstly discovered in the magnetar 1E 2259+586 (Archibald et al. 2013), it was proposed that anti-glitch is due to wind braking of the magnetar (Tong 2014).
The magnetic fields of magnetars may be twisted compared with that of normal pulsars (Thompson et al. 2002). This may account for the outburst activities of magnetars, i.e. decreasing X-ray luminosity, shrinking hot spot, decreasing torque etc (Beloborodov 2009; Pavan et al. 2009; Tong 2019; Tong \& Huang 2020).

Based on previous experiences on magnetar anti-glitch and magnetosphere, the following three questions of magnetar SGR 1935+2154 will be addressed: (1) What's its spin-down mechanism during the persistent state? (2) Can its anti-glitch be understood similarly to the first anti-glitch? (3) Why its pulsed radio emission only last about one month?

\section{Spin-down mechanism of SGR 1935+2154 during the persistent state}

The magnetar SGR 1935+2154 has a period of $P=3.245 \ \rm s$, and period derivative of $\dot{P}=1.43\times 10^{-11} \ \rm s \ s^{-1}$ (Israel et al. 2016). Assuming that the neutron star is spin-down by magnetic dipole radiation in vacuum, the surface magnetic field at the equatorial region is (i.e., the characteristic magnetic field):
\begin{equation}
  B_{\rm c} = 3.2\times 10^{19} \sqrt{P \dot{P}} = 2.2 \times 10^{14} \ \rm G.
\end{equation}
Note that the magnetic field at the polar region is two times higher for a dipole field (Lyne \& Graham-Smith 2012). The polar region magnetic field is more relevant to particle acceleration and radio emission (Ruderman \& Sutherland 1975). The rotational energy loss rate of SGR 1935+2154 is about: $\dot{E}_{\rm rot} = 1.7\times 10^{34} \ \rm erg \ s^{-1}$. The soft X-ray flux of SGR 1935+2154 is about: $3\times 10^{-12} \ \rm erg \ cm^{-2} \ s^{-1}$ (Younes et al. 2022). Assuming a distance of $10 \ \rm kpc$ (Zhou et al. 2020), the isotropic soft X-ray luminosity of SGR 1935+2154 is: $L_{\rm x} = 3.6\times 10^{34} \ \rm erg \ cm^{-2} \ s^{-1}$. Therefore, the X-ray luminosity is slightly higher than the rotational energy loss rate.

The magnetic dipole braking assumption does not consider the effect of the magnetosphere. In the case of magnetars, its X-ray emission is believed to come from the magnetic energy release. The origin for its radio emission is not certain at present. The magnetic energy output dominates over that of the rotational energy loss rate. Therefore, most of the rotational energy and angular momentum of the central neutron star is carried away by the particle wind (which is due to magnetic energy release). This is the idea of wind braking of magnetars (Harding et al. 1999; Tong et al. 2013). According to the wind braking model, the rotational energy loss rate is enhanced due to the particle wind (Harding et al. 1999; Tong et al. 2013):
\begin{equation}\label{eqn_wind_braking}
  \dot{E}_{\rm w} = \dot{E}_{\rm d} \left( \frac{L_{\rm p}}{\dot{E}_{\rm d}} \right)^{1/2},
\end{equation}
where $\dot{E}_{\rm w}$ is the rotational loss rate due to the particle wind, $\dot{E}_{\rm  d}$ is the rotational energy loss rate due to magnetic dipole braking, and $L_{\rm p}$ is the particle wind luminosity. The rotational energy loss rate is enhanced for a strong particle wind $L_{\rm p} \gg \dot{E}_{\rm d}$. The magnetic field in the strong wind case will be much lower than characteristic magnetic field (Tong et al. 2013). When the particle wind luminosity is comparable with the dipole rotational energy loss rate $L_{\rm p} \sim \dot{E}_{\rm d}$, the wind braking rotational energy loss rate is similar to the magnetic dipole braking case. The magnetic field may be approximated by the characteristic magnetic field.

Since both the particle wind and X-ray luminosity are powered by the magnetar's magnetic energy release, it is possible that: $L_{\rm p} \sim L_{\rm x}$ (Tong et al. 2013). For most magnetars, their X-ray luminosity are much higher than their rotational energy loss rate (i.e., the name ``magnetar"), therefore the true magnetic field of magnetars may be much lower than the characteristic magnetic field.
For SGR 1935+2154, its X-ray luminosity is slightly higher than its rotational energy loss rate. Therefore, the existence of a particle wind will only slightly enhance the rotational energy loss rate compared with the magnetic dipole case. Therefore, the true magnetic field of SGR 1935+2154 may be approximated by its characteristic magnetic field. The spin-down mechanism of SGR 1935+2154 may be approximated by the magnetic dipole braking in the persistent state.

\section{Anti-glitch of SGR 1935+2154 in the wind braking scenario}

After the discovery of the first anti-glitch (Archibald et al. 2013), it was proposed that: there is no anti-glitches, and anti-glitch is just a period of enhanced spin-down in the wind braking model. Due to the sparse of observations, a discrete decrease of frequency is seen, i.e., anti-glitch\footnote{In the following, the name ``anti-glitch"  will still be used to  denote the corresponding timing event in the magnetar.} (Tong 2014).

For a net decrease of frequency: $|\Delta \nu| = 1.8\times 10^{-6} \ \rm Hz$ (Younes et al. 2022), the decrease of rotational energy is: $|\Delta E_{\rm rot}| = 2\times 10^{40} \ \rm erg$. In the wind braking model, the corresponding rotational energy is carried away by the particle wind (Tong 2014, can also be deduced from eq.(\ref{eqn_wind_braking})):
\begin{equation}
 |\Delta E_{\rm rot}| = \dot{E}_{\rm d}^{1/2} L_{\rm p}^{1/2} \Delta t = \dot{E}_{\rm d}^{1/2} \Delta E_{\rm w}^{1/2} \Delta t^{1/2},
\end{equation}
where $\Delta t$ is the time duration for the particle wind, and $\Delta E_{\rm w}$ is total energy of the particle wind. The anti-glitch of SGR 1935+2154 occurred between October 1st and October 6th (Younes et al. 2022), therefore the maximum possible time duration is about 5 days. Similar to the calculation for 1E 2259+586 (Tong 2014), the particle wind energy for the anti-glitch in SGR 1935+2154 is:
\begin{equation}
 \Delta E_{\rm w} =8\times 10^{40} I_{45}^2 b_0^{-2} \left( \frac{|\Delta \nu|}{1.8\times 10^{-6} \,\rm Hz} \right)^2
		    \left( \frac{5 \,\rm days}{\Delta t} \right) \,\rm erg,
\end{equation}
where $b_0$ is the star's polar magnetic field in units of $4\times 10^{14} \ \rm G$ (the value by assuming magnetic dipole braking). The corresponding  wind luminosity is:
\begin{equation}
 L_{\rm p} =1.8\times 10^{35} I_{45}^2 b_0^{-2} \left( \frac{|\Delta \nu|}{1.8\times 10^{-6} \,\rm Hz} \right)^2
		    \left( \frac{5 \,\rm days}{\Delta t} \right)^2 \,\rm erg \,s^{-1}.
\end{equation}
Since the time duration is: $\Delta t \le 5 \ \rm days$, the particle wind energy and luminosity here are lower limits.
The particle wind luminosity is about five times higher than the magnetar's persistent X-ray luminosity. Therefore, it is may be easy to achieve. A slightly higher particle luminosity is also consistent with the non-detection of significant X-ray variations in SGR 1935+2154 (Younes et al. 2022).

In the discovery paper (Younes et al. 2022), the authors also discussed the possibility of wind braking for the anti-glitch. However, (1) they assumed mass loading of the particle wind. The wind from normal pulsars may come from pair production in open field line regions. The wind of magnetars may also come from pair production in the open or closed field line regions (Beloborodov 2009; Tong 2019). There is no need for  mass loading of the particle wind. (2) They implicitly assume that anti-glitch is a short duration event, like that of the glitches. Therefore, the authors try to constrain the time for the anti-glitch event observationally. However, in the wind braking model anti-glitch is an enhanced spin-down process, not an instantaneous event.

A particle wind lasting for some time interval (Tong 2014) or a corona-mass-ejection-like event (Lyutikov 2013) can both account for the anti-glitches. The anti-glitch in SGR 1935+2154 is accompanied by radio activities. This is consistent with the magnetospheric origin for anti-glitches (also noted in the discovery paper). Furthermore, no magnetar burst is seen around the anti-glitch epoch (Younes et al. 2022). The absence of burst may be understood  more naturally in the wind braking model.

\section{Transient pulsed radio emission of SGR 1935+2154}

\begin{figure}
  \centering
  \includegraphics[width=0.45\textwidth]{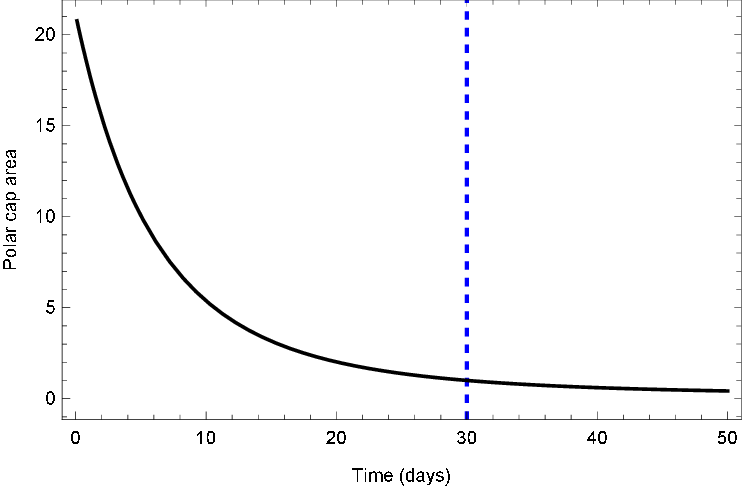}\\
  \caption{Evolution of the polar cap area with time. The polar cap area here is dimensionless. It is normalized to the polar cap area at $t=30 \ \rm days$. At a given emission height, the radio emission beam area is proportional to the polar cap area. Therefore, this curve is also the evolution of radio emission beam with time. The typical duration of transient pulsed radio emission in SGR 1935+2154 is about 30 days.}\label{fig_polar_cap}
\end{figure}

Transient pulsed radio emission is detected lasting for about 30 days after the epoch of anti-glitch (Younes et al. 2022; Zhu et al. 2020). This makes SGR 1935+2154 the sixth magnetar to have pulsed radio emissions (Camilo et al. 2006; Huang et al. 2021). The coherent radio emission mechanism of pulsars, magnetars and fast radio bursts is still uncertain at present (Lyne \& Graham-Smith 2012). It may be easier to answer why the radio emission of SGR 1935+2154 only turns on for about 30 days.

Magnetars may have twisted magnetospheres compared with that of normal pulsars (Thompson et al. 2002). The untwisting of the magnetic field may be responsible for the X-ray outburst of magnetars (Beloborodov 2009; Pavan et al. 2009; Tong 2019; Tong \& Huang 2020). Due to the twist of magnetic field, magnetars may have larger polar cap than the dipole magnetic field case (Tong 2019). During the untwisting process, the polar cap shrinks with time, which may corresponds to the shrinking hot spot of magnetars X-ray observations (Tong 2019; Tong \& Huang 2020). At the same time, the beam of the radio emission will also be proportional to the polar cap area at a given emission height (Lyne \& Manchester 1988; Rankin 1993; Wang et al. 2013). A time varying emission beam may account for the nulling and intermittence of pulsars (Timokhin 2010; Huang et al. 2016). Similarly, a decreasing emission beam during outburst may also explain the transient behavior of magnetar pulsed radio emissions.

According to the analytical treatment of magnetar outburst (Tong \& Huang 2020, section 5 there), for a self-similar globally twisted dipole magnetic field (Wolfson 1995; Thompson et al. 2002; Pavan et al. 2009; Tong 2019), the polar cap opening angle is:
\begin{equation}\label{eqn_thetapc}
  \theta_{\rm pc} = \left ( \frac{R}{R_{\rm lc}} \right)^{n/2},
\end{equation}
where $R$ is the neutron star radius, $R_{\rm lc}$ is the light cylinder radius, and $n$ is the parameter characterising the twist of the magnetic field. For a twisted magnetic field, the magnetic field varies with radius as: $B(r) \propto r^{-(2+n)}$ (Wolfson 1995). It can be seen that $n=1$ corresponds to the dipole case, $n=0$ cooresponds to the split monopole case, and $0<n<1$ corresponds to the twisted dipole case. From eq.(\ref{eqn_thetapc}), it can be seen that magnetars with twisted magnetic field will have larger polar caps compared with the dipole case (Tong 2019). By introducing a typical decaying timescale of the magnetic energy, the twist of the magnetic field evolves with time as (Tong \& Huang 2020):
\begin{equation}
  n(t) = 1- (1-n_0) e^{-t/\tau},
\end{equation}
where $n_0$ is the initial value of $n$ (which may be related to the hot spot radius), and $\tau$ is the typical decaying timescale of the magnetic energy (which may be determined by the particle acceleration process in the twisted magnetosphere, Beloborodov 2009; Tong 2019). For an initial value of $n_0=0.5$ (which corresponds to a polar cap about $1 \ \rm km$), and $\tau= 30 \ \rm days$ (which is about the duration of the transient radio emission), the polar cap area evolution with time is shown in figure \ref{fig_polar_cap}. It can be seen that the polar cap area decreased by a factor of 20. At a given emission height, the radio emission beam area is proportional to the polar cap area. Then the radio emission beam will also be smaller by a factor of 20 during the magnetar outburst. The pulsed radio emission is turned on when the enlarged emission beam encloses the line of sight. When the emission beam decreased significantly, the line of sight is out of the emission beam. The radio emission will be ``turned off".

Here the appearance and disappearance of pulsed radio emission is mainly discussed. The fast radio bursts from SGR 1935+2154 need independent studies. They may be similar to the magnetar giant flares (Lyutikov 2006; Yu 2012; Yu \& Huang 2013 in the magnetar context; Popov \& Postnov 2007; Katz 2016; Beloborodov 2017; Margalit \& Metzger 2018 in the fast radio burst context). Similar things in the case of magnetars are: modeling bursts and giant flares is different from that of outburst and persistent emissions.

\section{Discussions: other possibilities that are less likely}

\begin{figure}
  \centering
  \includegraphics[width=0.45\textwidth]{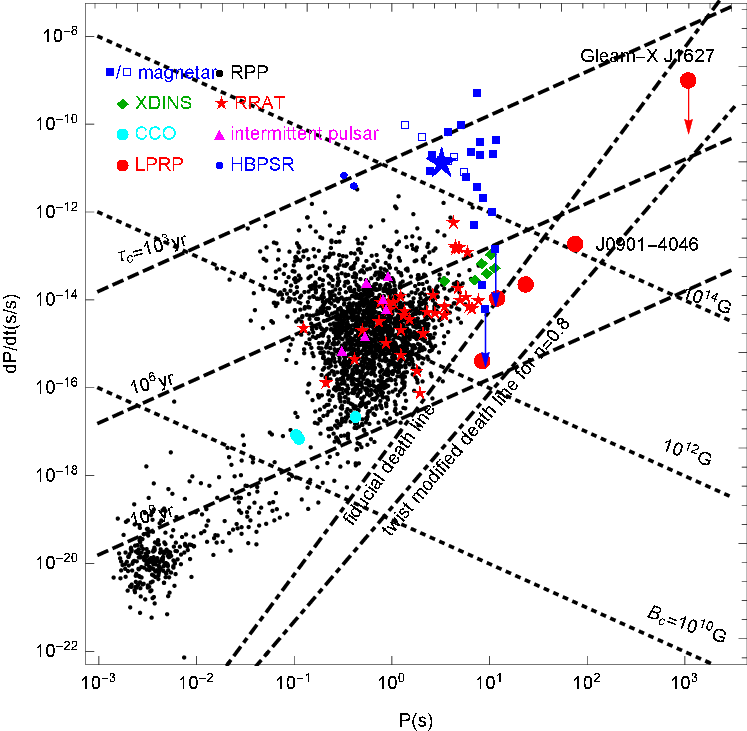}\\
  \caption{Radio pulsar death line on the $P$-$\dot{P}$ diagram. Both the fiducial death line and the death line for a twisted dipole field (with $n=0.8$) are shown. For $n=0.5$, the position of the death line will be even lower. The blue star is the magnetar SGR 1935+2154. Adapted from figure 2 in Tong (2022). The labeling of different pulsar like objects can be found in figure 1 in Tong \& Huang (2020).}\label{fig_PPdot}
\end{figure}

Although magnetars generally have a rotational period about several seconds, most of them still lie above the radio pulsar death line (see figure \ref{fig_PPdot}). This is because their dipole magnetic field can be as high as $10^{14}- 10^{15} \ \rm G$. If the particle wind luminosity is as high as $10^{38} \ \rm erg \ s^{-1}$ (i.e., $10^4$ times higher than the rotational energy los rate), then the required magnetic field will be $100$ times lower (see eq.(\ref{eqn_wind_braking})). Furthermore, if the magnetar's spin-down torque is dominated by a fallback disk (Chatterjee et al. 2000; Alpar 2001; Benli \& Ertan 2016), the star's true magnetic field may be in the range $10^{12}-10^{13} \ \rm G$. With such a normal magnetic field strength, for a period about $10 \ \rm s$, the star may already lie below the death line. During the outburst, a larger polar cap is obtained due to the twist of the magnetic field (see eq.(\ref{eqn_thetapc})). This will result in a higher acceleration potential across the polar cap and may explain the radio turn-on of SGR 1935+2154. During the outburst decaying phase, a shrinking polar cap will result in a decreasing acceleration potential. When the acceleration potential decreases below a threshold value, the radio emission may cease to operate.

\begin{figure}
  \centering
  \includegraphics[width=0.45\textwidth]{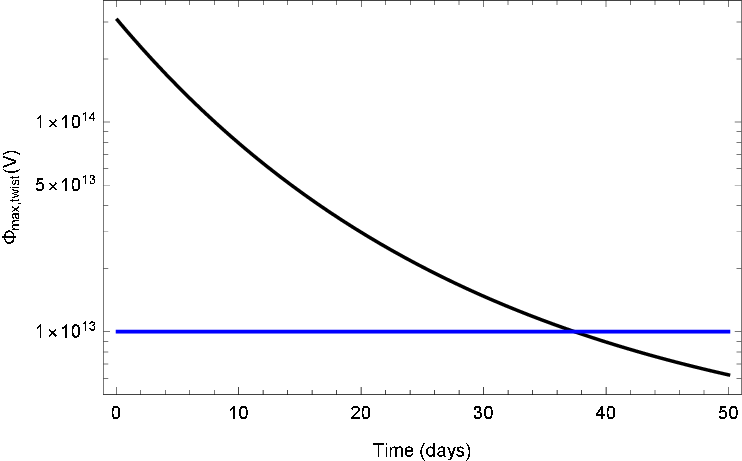}\\
  \caption{Evolution of the maximum acceleration potential with time (black line), for a twisted magnetosphere. The horizontal blue line is the threshold potential for the turn-on of radio emissions. }\label{fig_Phimaxtwist}
\end{figure}

A fiducial definition of the radio pulsar death line is that the maximum acceleration potential across the polar cap is about $10^{12} \ \rm V$ (Ruderman \& Sutherland 1975; Zhou et al. 2017). The maximum acceleration potential is (Ruderman \& Sutherland 1975; Zhou et al. 2017):
\begin{equation}
  \Phi_{\rm max} = \frac{B_{\rm p} R^2 \Omega}{2c} \sin^2\theta_{\rm pc},
\end{equation}
where $B_{\rm p}$ is the surface magnetic field at the magnetic pole region. For a given magnetic field strength $4 \times 10^{12} \ \rm G$ (one hundred times smaller than the characteristic magnetic field of SGR 1935+2154), employing a decreasing polar cap (eq.(\ref{eqn_thetapc})), the evolution of acceleration potential is shown in figure \ref{fig_Phimaxtwist}. In the figure, a higher threshold acceleration potential of $10^{13} \ \rm V$ is shown, due to uncertainties in the definitions of pulsar death line (Chen \& Ruderman 1993; Zhang et al. 2000) or it can be viewed as fine tuning of parameters. When the maximum acceleration potential drops below the threshold value, the star can no longer sustain the required potential of pair production for coherent radio emissions. The position of the death line  is also lower on the $P$-$\dot{P}$ diagram of pulsars, shown in figure \ref{fig_PPdot}. As can be seen from the figure, most of the magnetars lie already above the fiducial death line. A lower death line for a twisted magnetic  field only strengthens this point.

This possibility (decreasing acceleration potential) may be less likely compared with the previous one (decreasing emission beam). The reason is that additional assumptions are required to make the magnetar have  much lower dipole field and lie below the death line in the persistent state. In the case of decreasing emission beam, the SGR 1935+2154 already lies above the pulsar death line with a magnetar strength magnetic field. An even higher acceleration potential during the outburst state may help to generate the pulsed radio emissions. But the acceleration potential is not the dominate factor in this case. A similar point is that some pulsars are only detected at high energy band (other than radio) (Manchester et al. 2005). This may because their radio emission beam does not point toward us, not because they are radio quiet.

\section{conclusions}

The triplet of an anti-glitch, fast radio bursts and pulsed radio emission in SGR 1935+2154 can unveil the physics of the magnetar magnetosphere:
\begin{enumerate}
  \item The magnetar does not have a strong particle wind during its persistent state. Its  spin-down mechanism can be approximated by the magnetic dipole braking in the persistent state.
  \item The anti-glitch may be due to an enhanced particle wind, similar  to the first anti-glitch. The anti-glitch in SGR 1935+2154 is consistent with the wind braking model of anti-glitches (Tong 2014).
  \item The transient pulsed radio emission may be due to a decreasing emission beam after outburst.

  \item An enhanced acceleration potential and pulsar death line may not be the dominate factor.
\end{enumerate}
The mechanism of fast radio burst needs independent studies.

\section*{acknowledgments}
This work is supported by National SKA Program of China (No. 2020SKA0120300) and NSFC (12133004).




\label{lastpage}


\begin{thebibliography}{99}

\bibitem{Archibald2013}
Archibald, R. F., Kaspi, V. M., Ng, C. Y., et al. 2013 Nature, 497, 591

\bibitem{Alpar2001}
Alpar, M. A. 2001, ApJ, 554, 1245

\bibitem{Beloborodov2009}
Beloborodov, A. M. 2009, ApJ, 703, 1044

\bibitem{Beloborodov2017}
Beloborodov, A. M. 2017, ApJL, 843, L26

\bibitem{Benli2017}
Benli, O., \& Ertan, U. 2016, MNRAS, 457, 4114

\bibitem{Bochenek2020}
Bochenek, C. D., Ravi, V., Belov, K. V., et al. 2020, Nature, 587, 59

\bibitem{Camilo2006}
Camilo, F., Ransom, S. M., Halpern, J. P., et al. 2006, Nature, 442, 892

\bibitem{Chatterjee2000}
Chatterjee, P., Hernquist, L., \& Narayan, R. 2000, ApJ, 534, 373

\bibitem{Chen1993}
Chen K., \& Ruderman M. 1993, ApJ, 402, 264

\bibitem{CHIME2020}
CHIME/FRB Collaboration: Andersen, B. C., Bandura, K., Bhardwaj, M., et al., 2020, Nature, 587, 54

\bibitem{Good2020}
Good, D. \& CHIME/FRB Collaboration 2020, The Astronomer's Telegram 14074, 1

\bibitem{Harding1999}
Harding, A. K., Contopoulos, I., \& Kazanas, D. 1999, ApJL, 525, L125

\bibitem{Huang2016}
Huang, L., Yu, C., \& Tong, H. 2016, ApJ, 827, 80

\bibitem{Huang2021}
Huang, Z. P., Yan, Z., Shen, Z. Q., et al. 2021, MNRAS, 505, 1311

\bibitem{Israel2016}
Israel, G. L., Esposito, P., Rea, N., et al. 2016, MNRAS, 457, 3448

\bibitem{Katz2016}
Katz, J. I. 2016, ApJ, 826, 226

\bibitem{Lyne2012}
Lyne, A. G., \& Graham-Smith, F. 2012, Pulsar astronomy (4th ed.), Cambridge University Press, Cambridge

\bibitem{Lyne1988}
Lyne A. G., \& Manchester R. N. 1988, MNRAS, 234, 477

\bibitem{Lyutikov2006}
Lyutikov, M. 2006, MNRAS, 367, 1602

\bibitem{Lyutikov2013}
Lyutikov, M. 2013, arXiv:1306.2264

\bibitem{Manchester2005}
Manchester, R. N., Hobbs, G. B., Teoh, A., \& Hobbs, M. 2005, AJ, 129, 1993

\bibitem{Margalit2018}
Margalit B., \& Metzger B. M. 2018, ApJL, 868, L4


\bibitem{Pavan2009}
Pavan, L., Turolla, R., Zane, S., \& Nobili, L. 2009, MNRAS, 395, 753

\bibitem{Popov2007}
Popov, S. B., \& Postnov, K. A. 2007, arXiv:0710.2006

\bibitem{Rankin1993}
Rankin, J. M. 1993, ApJ, 405, 285

\bibitem{Ruderman1975}
Ruderman, M. A., \& Sutherland, P. G. 1975, ApJ, 196, 51

\bibitem{Thompson2002}
Thompson, C., Lyutikov, M., \& Kulkarni, S. R. 2002, ApJ, 574, 332

\bibitem{Timokhin2010}
Timokhin, A. N. 2010, MNRAS, 408, L41

\bibitem{Tong2013}
Tong, H., Xu, R. X., Song, L. M., \& Qiao, G. J. 2013, ApJ, 768,144

\bibitem{Tong2014}
Tong, H. 2014, ApJ, 784, 86


\bibitem{Tong2019}
Tong, H. 2019, MNRAS, 489, 3769

\bibitem{Tong2020}
Tong, H., \& Huang, L. 2020, MNRAS, 497, 2680

\bibitem{Tong2022}
Tong, H. 2022, arXiv:2204.01957


\bibitem{Wang2013}
Wang P. F., Han J. L., \& Wang C. 2013, ApJ, 768, 114

\bibitem{Wolfson1995}
Wolfson, R. 1995, ApJ, 443, 810

\bibitem{Younes2022}
Younes, G., Baring, M. G., Harding, A. K., et al. 2022, arXiv:2210.11518

\bibitem{Yu2012}
Yu, C. 2012, ApJ, 757, 67

\bibitem{Yu2013}
Yu, C., \& Huang, L. 2013, ApJL, 771, L46

\bibitem{Zhang2000}
Zhang, B., Harding, A. K., \& Muslimov, A. G. 2000, ApJL, 531, L135

\bibitem{Zhou2017}
Zhou, X., Tong, H., Zhu, C., \& Wang, N. 2017, MNRAS, 472, 2403

\bibitem{Zhou2017}
Zhou, P., Zhou, X., Chen, Y., et al. 2020, ApJ, 905, 99

\bibitem{Zhu2020}
Zhu, W. W., et al. 2020, The Astronomer's Telegram 14084, 1

\end{thebibliography}
\end{document}